# Impact of guided reflection with peers on the development of effective problem solving strategies and physics learning

Andrew J. Mason[1] and Chandralekha Singh[2]

Department of Physics and Astronomy, University of Central Arkansas[1], University of Pittsburgh[2]

## Introduction

Students must learn effective problem solving strategies in order to develop expertise in physics.[1-6] Effective problem solving strategies include a conceptual analysis of the problem followed by planning of the solution, and then implementation, evaluation and reflection upon the process.[1-6] Research suggests that converting a problem from the initial verbal representation to other suitable representation, e.g., diagrammatic representation, during the initial conceptual analysis can facilitate further analysis of the problem.[6] But without guidance, many introductory physics students solve problems using superficial clues and cues and do not perceive problem solving as an opportunity for learning.[1-6] Here, we describe a study which suggests that engaging students in reflection with peers about effective problem solving strategies while effective approaches are modeled for them and prompt feedback is provided may enhance desirable skills.[7]

According to the field-tested cognitive apprenticeship model, students can learn effective problem solving strategies if the instructional design involves three essential components: modeling, coaching & scaffolding, and weaning.[8] In this approach, "modeling" means that the instructor demonstrates and exemplifies the skills that students should learn (e.g., how to solve physics problems systematically). "Coaching & scaffolding" means that students receive appropriate guidance and support as they actively engage in learning the skills necessary for good performance. "Weaning" means reducing the support and feedback gradually to help students develop self-reliance.

In traditional physics instruction, especially at the college level, there is often a lack of coaching and scaffolding.[9,10] The situation is often akin to a piano instructor demonstrating for the students how to play the piano and then asking them to go home and practice it - the lack of opportunity for prompt feedback and scaffolding can be detrimental to learning. One activity that may help students learn effective problem solving strategies, while simultaneously learning physics content, is reflection with peers. In this approach, students reflect not only on their own solutions to problems, but also upon their peers' solutions. Peer collaboration as a learning tool has been exploited in many instructional settings and with different types and levels of student populations.[7,11-13] Although the details vary, students can learn from each other in many different environments.

Integration of peer interaction with traditional lectures has been popularized in the physics community by Mazur.[11] In Mazur's approach, the instructor poses conceptual problems in the form of multiple-choice questions to students throughout the lecture. On one hand, students learn about the level of understanding that is desired by the instructor by discussing the concrete questions posed with each other. The feedback obtained by the instructor is also valuable because the instructor measures the fraction of the class that has understood the concepts at the desired level. This peer-instruction strategy keeps students alert during lectures and helps them monitor their learning, because not only do students have to answer the questions, they must explain their answers to their peers. In addition, Heller et al. have shown that collaborative problem solving in the context of quantitative "context-rich" problems is valuable both for learning physics and for developing effective problem solving strategies.[12] Our prior research has shown that even with minimal guidance from the instructors, students can benefit from peer interaction.[13] In particular, those who worked with peers not only outperformed an equivalent group of students who worked alone on the same task, but collaboration with a peer led to "co-construction" of knowledge in 29% of the cases.[12] Co-construction of knowledge occurs when neither student who engaged in the peer collaboration was able to answer the questions before the collaboration, but both were able to answer them after working with a peer on a post-test given individually to each person.[12]

## Study Design

In this study, the recitation sections for 200 students in an algebra-based college introductory physics class at a typical state university were broken into the "Peer Reflection" (PR) group and the traditional group. The traditional group (daytime section) had 107 students in three recitation sections, and the PR group (evening section) had 93 students in two recitation sections. While the traditional lectures and all out of class assignments for all recitation sections were identical to the best of the course instructor's ability, the recitations for the traditional group and PR group were structured differently. In the traditional group recitations, the TA would solve selected assigned homework problems on the blackboard and field questions from students about their homework before assigning a quiz in the last 20 minutes of the recitation class. The quiz for the traditional group typically had three problems per week based upon the homework that was assigned for that week. For fairness, the cumulative scores in the course for the PR group and the traditional group were curved separately for determining the final course grades.

In addition, each week, regardless of the recitation group they belonged to, all students were supposed to enter answers to the assigned homework problems (based upon the material covered in the previous week) using an online homework system for some course credit. Moreover, students were supposed to submit a paper copy of the homework problems which included the details of the problem solving approach at the end of the recitation class to the TA for partial course credit. While the online homework solution was graded for correctness, the TA only graded the paper copies of the submitted homework for completeness.

The peer-reflection process in the PR group focuses on the coaching and scaffolding aspect of the cognitive apprenticeship model and requires students to evaluate their solutions and those of their peers during each recitation of the semester. The students in the PR group first worked alone on the problems as part of the homework and then brainstormed together to figure out the strengths and weaknesses of the solutions to the problems generated by each student as they tried to converge on the best solutions. The specific strategy used by the students in the PR group involved reflection upon their problem solving approaches with their peers in the recitations, while the TA and the undergraduate teaching assistants (UTAs) exemplified the effective problem solving heuristics. We note that the course instructor met with the TA and UTAs each week to ensure that they knew how to model effective problem solving approaches, provide useful feedback to students and scaffold their learning.

Students in the PR group, who reflected in small teams on selected problems from the homework, discussed why solutions of some students employed better problem solving strategies than others. Each small team in the PR group discussed which student's homework solutions employed the most effective problem solving heuristics and used them as the basis to select a "winner". Then, three teams combined into a larger team and repeated the process of determining the "winning" solution (see Figure 1). Typically, once three "finalists" were identified in this manner, the TA and UTAs put each finalist's solution on a projector and discussed what they perceived to be good problem-solving strategies used in each solution and what can be improved. After the active engagement with peers, the opportunity students got to learn from the TA and UTAs about their critique of each solution highlighting the strengths and weaknesses was valuable for coaching and scaffolding. For example, TAs and UTAs emphasized and praised good diagrams and clear procedures and appropriate concepts being applied before implementing the plan. The TAs and UTA' explicit open critiques were inspired by the expertise literature which documents that often experts' knowledge remains tacit and needs to be made explicit.[1-2] In particular, what the TAs and UTAs attempted to do was to make expert thinking more visible or explicit to novices to learn from and emulate.

Finally, each student used clickers to vote on the best overall solution with regard to the problem solving strategies used. There was a reward system related to course credit that encouraged students to select the solution with the best problem solving strategy in each round (for the winning solution, all students who were part of judging whether that solution had the best problem solving approach out of all student solutions in that group at any stage of judging were given bonus points). We note that the traditional group recitation quiz questions were based upon the homework questions selected for "peer reflection" in the PR group recitations.

**Figure 1.** Illustration of the team structure at the three stages of peer-reflection activities. Before students voted in the third round using the clickers, the TA and UTAs critiqued each of the three solutions at the final stage. Due to lack of space, only 3 team members per team are shown in round 1 but there were on average five members in each group in this round. The consolation prize winners in gray obtained 1/3rd of the course credit awarded to the winner.

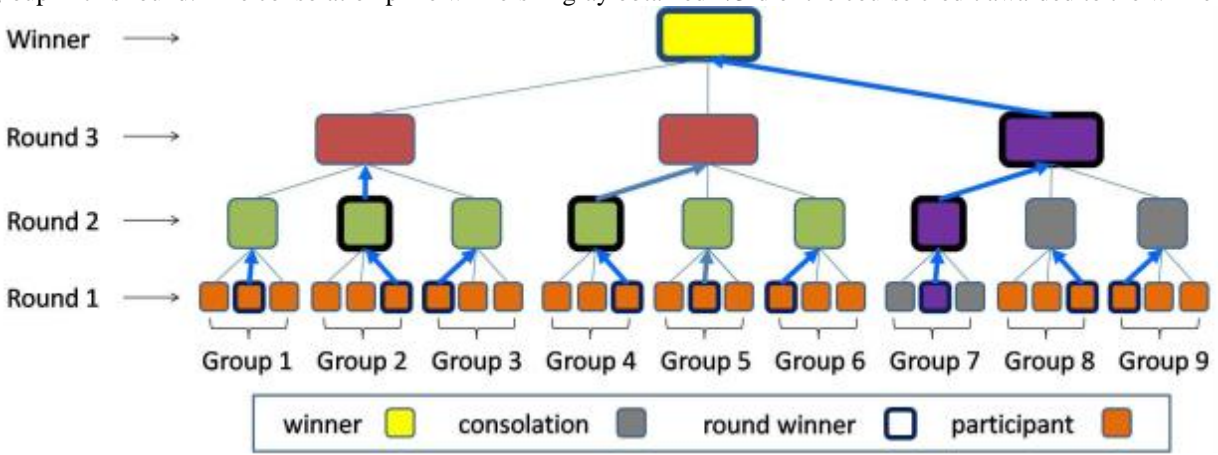

## Goals and Assessment Method

Our goals were to investigate whether students in the PR group use better problem-solving strategies, such as drawing diagrams, than students in the traditional group on the final exam at the end of the semester, and also whether there are differences in the overall performance of the two groups. We also explored whether students who perform well are the ones who are more likely to draw diagrams or write scratch work, even when there is no partial credit for these activities. We note that drawing a diagram was singled out as crucial because it is an effective problem solving strategy that can easily be quantified objectively as opposed to other strategies such as analyzing the problems qualitatively in terms of what concepts to apply or applying general procedures.

The final exam had 40 multiple-choice questions, half of which were quantitative and half were conceptual. For the multiple-choice questions, students do not have to show their work to arrive at the answer. However, students may use effective approaches to problem solving, such as drawing a diagram or writing down their plan, if they think it may help them answer a question correctly. One issue we investigated is whether the students considered the diagrams or the scratch work to be beneficial and used them while solving problems, even though they knew that no partial credit was given for showing work.

A multiple-choice exam can be a novel tool for assessment, especially if students are given opportunity to write in the exam booklet what they think is helpful for problem solving. It is helpful to observe students' problem solving strategies in a more "native" form closer to what they really think is helpful for problem solving, instead of what the instructor wants them to write down, or filling the page with irrelevant equations and concepts with the hope of getting partial credit for the effort when a free-response question is assigned. While students knew that the only thing that counted for their grade was whether they chose the correct option for each multiple-choice question, each student was given an exam notebook which he/she could use for scratch work. We hypothesized that even if the final exam questions were in the multiple-choice format, students who value effective problem solving strategies will take the time to draw more diagrams and do more scratch work even if there was no course credit for such activities.

Our assessment method[7] involved counting the number of problems with diagrams and scratch work. We decided to count any comprehensible work done in the exam notebook other than a diagram as a scratch work. Instead of using subjectivity in deciding how "good" the diagrams or scratch work for each student for each of the 40 questions were, we only counted the number of problems with diagrams drawn and scratch work done by each student. For example, if a student drew diagrams for 7 questions out of 40 questions and did scratch work for 10 questions out of 40 questions, we counted it as 7 diagrams and 10 instances of scratch work. However, we will later comment on a potential indirect measure of the average quality of the diagrams or the scratch work from students in the PR and traditional groups.

## Results

We note that there is some evidence from the previous year in which the same instructor taught this course that the evening section was somewhat weaker and did not perform as well overall as the daytime section (in the previous year, the midterm and final exam averages for the daytime section were respectively 72.0% and 55.7%, vs. 65.8% and 52.7% for the evening section). The following year in which the daytime section was the traditional group, the midterm and final exam averages were respectively 78.8% and 58.1%, vs. 74.3% and 57.8% for the evening section (which was the peer reflection group). Thus, there is essentially no difference between the final exam score of the traditional and PR groups even though the PR group was likely to be the weaker evening section with lower midterm exam performance.

Moreover, on the final exam with only multiple-choice questions, the PR group drew diagrams on significantly more problems than the traditional group although there is no difference between the number of problems with scratch work (see Table 1). Since there was no partial credit for drawing the diagrams in the scratch books, students did not draw diagrams simply to get credit for the effort shown and must value the use of diagrams for solving problems if they drew them. Moreover, the improvement in the averaged student performance from the midterm to the final exam was larger for the PR group than for the traditional group. In addition, we find that diagrams helped with problem solving and students who drew diagrams for more problems performed better than others regardless of whether they belonged to the traditional group or the PR group.

**Table 1.** Comparison of the average number of problems per student with diagrams and scratch work by the traditional group and the PR group in the final exam. Statistical analysis shows that the PR group has significantly more problems with diagrams than the traditional group. The average number of problems with scratch work per student in the two groups is not significantly different. There are more quantitative problems with diagrams drawn and scratch work done than conceptual problems.

| Question type | Number of problems with diagrams | | | Number of problems with scratch work | | |
| --- | --- | --- | --- | --- | --- | --- |
| | All questions | Quantitative | Conceptual | All questions | Quantitative | Conceptual |
| Traditional | 7.0 | 4.3 | 2.7 | 20.2 | 16.0 | 4.2 |
| PR | 8.6 | 5.1 | 3.5 | 19.6 | 15.6 | 4.0 |

In summary, our findings can be broadly classified into inter-group and group-independent categories. The inter-group findings that show the difference between the traditional group and PR group can be summarized as follows:
- On the multiple-choice final exam, the PR group drew diagrams for more problems.
- The average improvement from the midterm to the final exam was larger for the PR group than for the traditional group. Also, the correlation between the number of problems with diagrams vs. the final exam score is somewhat higher for the PR group compared to the traditional group. Thus, while we did not distinguish between the quality of the diagrams or the scratch work, one possible interpretation of the stronger correlation for the PR group compared to the traditional group is that the students in the PR group on average drew more meaningful diagrams or did more meaningful scratch work that were more likely to help them answer the questions correctly.

Findings that are independent of group (which hold even when the traditional group and PR group are not separated and all students are considered together) can be summarized as follows:
- There was a significant positive correlation between how often students did scratch work or drew diagrams and how well they performed on the final exam regardless of whether they were in the traditional group or the PR group. In particular, those who performed well in the multiple-choice final exam (in which there was no partial credit for showing work) were much more likely to draw diagrams than the other students. While one may assume that high-performing students will draw more diagrams even when there is no partial credit for it, no prior research that we know of has explicitly demonstrated a correlation between the number of "genuinely drawn" diagrams and student performance at any level of physics instruction.
- Students who drew more diagrams or did more scratch work were more likely to show improvement from the midterm exam to the final exam than the other students.
- For both conceptual and quantitative questions, drawing diagrams helped students.
- Students in both groups were more likely to draw diagrams or do scratch work for quantitative problems than for the conceptual questions. While more scratch work is expected on quantitative problems, it is not clear *a priori* that more diagrams will be drawn for the quantitative problems than for the conceptual questions. We hypothesize that this trend may depend upon the expertise of the individual students, explicit training in effective problem solving strategies and the difficulty of the problems.

## Summary and Conclusion

This study evaluated the impact of helping students learn effective problem solving strategies by having them reflect on their solutions with peers and receive feedback from TAs and UTAs, who modeled effective approaches such as importance of drawing a diagram, writing down clear procedures and contemplating applicable concepts before implementing the plan. Students in the PR group first worked alone on the problems as part of the homework and then brainstormed together during recitation to figure out the strengths and weaknesses of the solutions to a problem generated by each student as they tried to agree on the best solution. Reflection with peers gave all students an opportunity to focus on and explain the effective aspects of the problem solving strategies. The improvement in the average student performance from the midterm to the final exam was larger for the PR group than the traditional group and students in the PR group on average drew more diagrams than those in the traditional group. We also found that drawing more diagrams correlated positively with student performance.

It is worthwhile for instructors to promote effective approaches to problem solving in a manner similar to that emphasized for the PR group. Giving students the opportunity to communicate with peers similar to the PR group can help students monitor their own learning and improve their ability to articulate their scientific point of view. It can also be beneficial in helping students develop the ability to communicate their scientific approaches especially because students must discuss with each other their rationale for perceiving one student's solution as superior in terms of its problem solving strategies than another solution. Although it is sometimes argued by instructors that the introductory physics recitation quizzes are essential to keep students engaged in the learning process during the recitations, the PR group was not adversely affected by having the weekly quizzes for the traditional group replaced by the peer reflection activities. The mental engagement of students in the PR group throughout the recitation class may have more than compensated for the lack of quizzes. The students in the PR group were evaluating their peers' work along with their own, which requires a high level of mental processing. They were comparing problem solving strategies, such as how to do a conceptual analysis and planning of the solution, why drawing two separate diagrams may be better in certain cases (e.g., before and after a collision) than combining the information into one diagram, how to define and use symbols consistently, etc. In addition, after the active engagement with peers, students got an opportunity to learn from the TA and UTAs about their critique of each solution highlighting the strengths and weaknesses, which is central to modeling, coaching and scaffolding. If TAs and UTAs are not available, instructors adopting or adapting a model similar to the PR group intervention in their classroom can play the role of the TAs and UTAs.

Andrew J. Mason is an assistant professor at the University of Central Arkansas. He obtained his Ph.D. in physics education research from the University of Pittsburgh and was a postdoctoral fellow at the University of Minnesota.

Chandralekha Singh is a professor of physics and the Founding Director of the Discipline-based Science Education Research Center at the University of Pittsburgh. She obtained her Ph.D. in theoretical condensed matter physics from the University of California Santa Barbara and was a postdoctoral fellow at the University of Illinois Urbana Champaign. She is a Fellow of the American Association of Physics Teachers and American Physical Society (APS) and has served as the chair of the APS Forum on Education. She was the chair of the Editorial Board of Physical Review Special Topics: Physics Education Research from 2010-2013. For more information about her work, visit https://sites.google.com/site/professorsinghswebpage/